%% file: report.tex
\documentclass[10pt,dvips,a4paper]{article}
\usepackage{epsfig}
\include{commands}
\usepackage{amssymb}
\usepackage{amsmath}
\usepackage{lineno}
\usepackage{setspace}
\usepackage[numbers]{natbib}

\begin{document}

\renewcommand{\title}{Influence of aggregate size and volume fraction on shrinkage induced micro-cracking of concrete and mortar}

\begin{center} \begin{LARGE} \textbf{\title} \end{LARGE} \end{center}

\begin{center} Peter Grassl*$^{1}$, Hong S. Wong$^{2}$ and Nick R. Buenfeld$^{2}$\\
$^{1}$~Department of Civil Engineering, University of Glasgow, Glasgow, UK\\
$^{2}$~Department of Civil Engineering, Imperial College London, London, UK\\
* Corresponding author: grassl@civil.gla.ac.uk

Submitted to Cement and Concrete Research, 31th of October 2008
\end{center}

\section*{Abstract}
In this paper, the influence of aggregate size and volume fraction on shrinkage induced micro-cracking and permeability of concrete and mortar was investigated.
Nonlinear finite element analyses of model concrete and mortar specimens with regular and random aggregate arrangements were performed.
The aggregate diameter was varied between 2~and~16~mm. Furthermore, a range of volume fractions between 0.1 and 0.5 was studied. 
The nonlinear analyses were based on a 2D lattice approach in which aggregates were simplified as monosized cylindrical inclusions.
The analysis results were interpreted by means of crack length, crack width and change of permeability.
The results show that increasing aggregate diameter (at equal volume fraction) and decreasing volume fraction (at equal aggregate diameter), increases crack width and consequently greatly increases permeability.

Keywords: Microcracking (B); Interfacial transition zone (B); Transport properties (C); Shrinkage (C); Aggregate (D); Lattice modelling.

\section{Introduction}
Drying of cement based composites, such as concrete and mortar, induces cracking if shrinkage of the constituents is either internally or externally restrained.
For example, non-uniform drying leads to a moisture gradient, which results in non-uniform shrinkage of the specimen.
Surface regions shrink faster than the inner bulk material, which results in surface cracking \cite{HwaYou84, BazRaf82}.
Additionally, shrinkage might be restrained by aggregates within the composite \cite{Hob74}.
Aggregate-restrained shrinkage can lead to micro-cracking, which strongly influences the transport properties of the material \cite{JenCha96, BisMie02, WonZobBue08}.
However, the evolution of micro-cracks and their dependence on the size and volume fraction of aggregates is not fully understood yet.
In \cite{WonZobBue08} it was observed that permeability increases with increasing aggregate size at a constant aggregate fraction. This result is surprising, since an increase of the aggregate size at a constant aggregate fraction is usually accompanied by a decrease in the volume of interfacial transition zones (ITZs), which are known to be more porous than the cement paste.
One hypothesis is that an increase of aggregate diameter at constant aggregate fraction results in an increase of micro-crack width, which is closely related to permeability \cite{WitWanIwaGal80}.
The objective of this work was to establish whether this size effect really occurs.
This will undoubtedly enhance the understanding of the link between micro-structure and macro-property, in particular the effect of microcracking on mass transport, which is a critical aspect for predicting durability and service-life.
Shrinkage induced micro-cracking has been investigated recently with a lattice model \cite{SchKoeBre07}. However, to the authors' knowledge the influence of the size of the aggregates on micro-cracking has not been analysed before.

In the present work, shrinkage induced micro-cracking of concrete was analysed by means of the nonlinear finite element method.
In concrete, stress-free cracks form by a complex nonlinear fracture process, during which energy is dissipated in zones of finite size.
There are three main approaches to describe fracture process zones within the finite element framework.
Continuum approaches describe the evolution of cracks as zones of inelastic strains by means of higher-order constitutive models such as integral-type nonlocal models \cite{BazJir02,PijBaz87}.
In hybrid approaches, cracks are modelled as displacement discontinuities, which are embedded into the continuum description \cite{MoeBel02, Jir00b, CamOrt96}.
Finally, discrete approaches represent the nonlinear fracture process by means of the failure of discrete elements, such as trusses and beams \cite{SchMie92b, HerHanRou89}.
In recent years, one type of discrete lattice approach based on the Voronoi tessellation has been shown to be very suitable for fracture simulations \cite{BolSai98, BolSuk05}.
This lattice approach is robust and computationally efficient. 
With specially designed constitutive laws, the results obtained with this modelling approach are mesh-independent.
In the present study, this lattice approach was used in combination with a damage-plasticity constitutive model, which was designed to result in mesh-independent responses \cite{GraRem08}.

The present study is based on several simplifications.
Shrinkage is represented by an eigenstrain, which was uniformly applied to the cement matrix only.
This is representative of autogenous shrinkage, but does not fully represent transient non-uniform shrinkage due to moisture gradients which may lead to more cracking near the surface than in the centre of a concrete element.
Furthermore, the only inclusions considered were aggregates, which were embedded in a uniform cement paste. 
Micro-cracking due to other inclusions, e.g. unhydrated cement and calcium hydroxide crystals, was not considered.
Aggregates were assumed to be separated from the cement paste by interfacial transition zones (ITZs), which were modelled to be weaker and more brittle than the cement matrix \cite{HsuSla63}. 
Shrinkage of aggregates, which might have a considerable influence on the overall shrinkage of concrete, was not modelled.
However, the shrinkage strain applied to the cement paste can be interpreted as the difference of uniform shrinkage in cement paste and aggregates. 
The mechanical response of aggregates was assumed to be elastic.
Furthermore, the study was limited to two-dimensional plane stress analyses, in which aggregates are idealised as cylindrical inclusions of constant diameter.

\section{Experimental results} \label{sec:experiments}
This section summarises the findings from a recent experimental study by Wong et al. \cite{WonZobBue08} as background to and motivation for the modelling work presented in this paper. 
Concretes and mortars with a range of aggregate contents, w/c ratio, binder type, curing period and preconditioning temperature were tested for oxygen diffusivity, oxygen permeability and water sorptivity. 
Thames Valley gravel (5-12.7mm) and sand ($<$5mm) were used as coarse and fine aggregates respectively. 
The main objective of the study was to examine the relative influences of ITZ and microcracking induced by oven drying on different transport properties, in light of the inconsistencies in previous investigations on ITZ and mass transport.

\begin{figure}
\begin{center}
\begin{tabular}{cc}
\epsfig{file=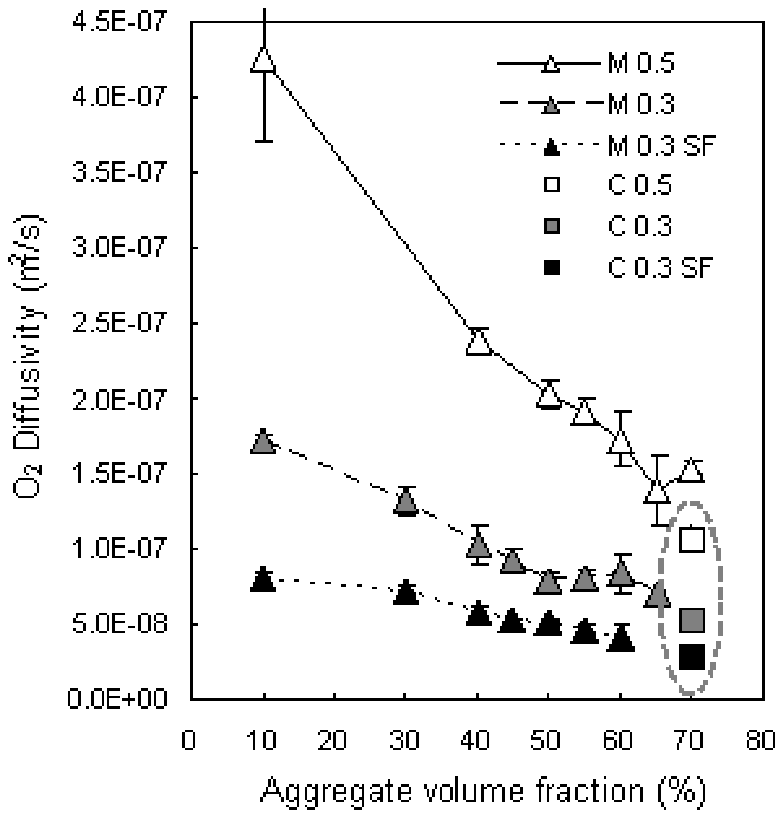,width=7cm} & \epsfig{file=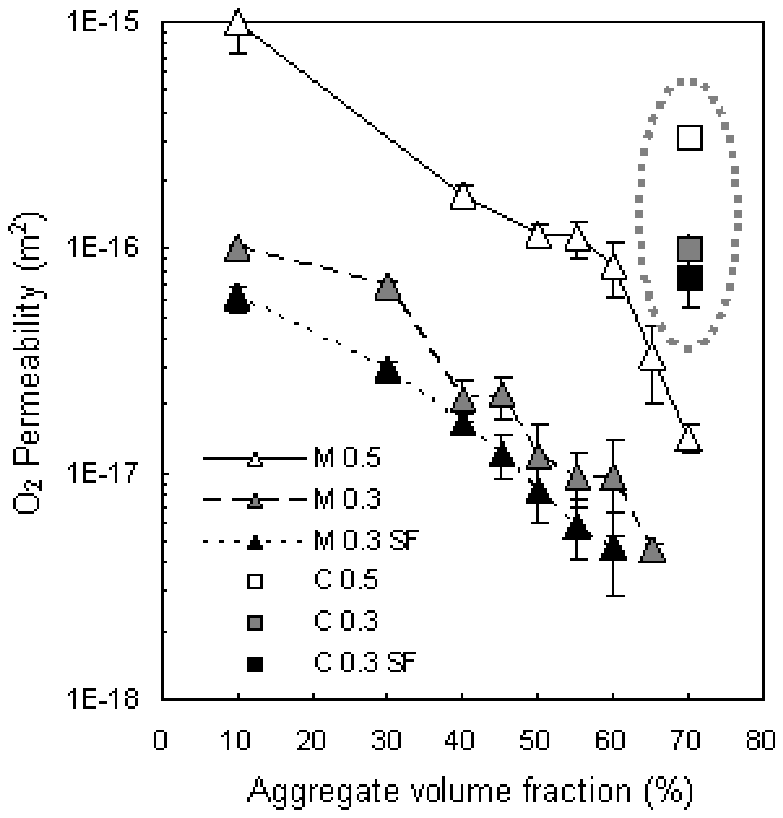,width=7cm} \\
(a) & (b)
\end{tabular}
\end{center}
\caption{Effect of aggregate volume fraction on a) oxygen diffusivity and b) oxygen permeability, of mortars and concretes with different w/c ratios (0.5 \& 0.3) and binder type. The samples were sealed cured in cling film for the first 90 days and then preconditioned at 50$^\circ$C, 10\% relative humidity for a further 90 days until constant mass prior to testing. The dotted circle highlights the anomalous trend for concrete samples, which are represented with squares. (M = mortar; C = concrete; SF = containing 8\% silica fume).}
\label{fig:exp}
\end{figure}

Fig.~\ref{fig:exp} reproduces the diffusivity and permeability results for 90-day cured samples that were preconditioned at 50$^\circ$C, 10\% r.h. 
Note that this represents only a small portion of the total results presented by Wong et al. \cite{WonZobBue08}, but captures the important aspects of their findings. 
Interested readers are referred to the original article for experimental details and in-depth discussion. 
It was observed that samples with low w/c ratio and containing silica fume recorded the lowest transport coefficients for all aggregate contents and curing ages, as expected. 
The transport coefficients also decreased with longer curing age, but increased at higher conditioning temperature. 
For mortars, the transport coefficients decreased significantly with increase in sand content. 
At very high sand fractions, the ITZs should be ‘overlapping’, but this did not have any detrimental effect on the transport properties. 
The trend remains consistent regardless of the type of transport property, w/c ratio, binder, curing age and even after severe oven drying treatment at 105$^\circ$C (results shown in \cite{WonZobBue08}). 

The most interesting finding was that concrete has roughly the same diffusivity and sorptivity as its analogous mortar, but significantly higher permeability (1-2 orders of magnitude), despite having about a third less ITZ (see Fig.~\ref{fig:exp}). 
This striking feature is also consistent across different w/c ratios, binder types, curing ages and conditioning regimes, increasing the confidence that it is not caused by experimental errors or a series of coincidences. 
Furthermore, the same sample was used for all three transport tests, and the permeability test was performed in between diffusivity and sorptivity, which discounts the possibility that the samples were damaged during permeability testing. 
However, the higher permeability in concretes cannot be attributed to the ITZs, because it would have a larger influence on the mortars if this was the case.

Wong et al. \cite{WonZobBue08} suggested that the higher permeability is due to more microcracks forming in concrete compared to in the analogous mortar, when both are subjected to the same drying condition. 
This is because aggregates in concrete have larger size distribution, which at the same volume fraction, equates to lower surface area to bond with the paste. 
Thus, any shrinkage (or expansion) in the paste would produce larger stresses at the interface and more microcracking compared to the analogous mortar. 
Since diffusivity and sorptivity are less sensitive to microcracking than permeability, these properties are not as affected. 
Microcracks in concretes also have lower specific lengths and tortuosity, because there are fewer obstructing aggregates per unit volume to arrest the cracks. Indeed, backscattered electron microscopy and image analysis on a few samples found that this may be the case. However, experimental results are often affected by many varying parameters that are difficult to isolate and control. Therefore, the present numerical study was carried out to confirm the trends observed in the experiments and to increase the understanding of the underlying mechanisms.

\section{Modelling approach}
\begin{figure}
\begin{center}
\begin{tabular}{cc}
\epsfig{file=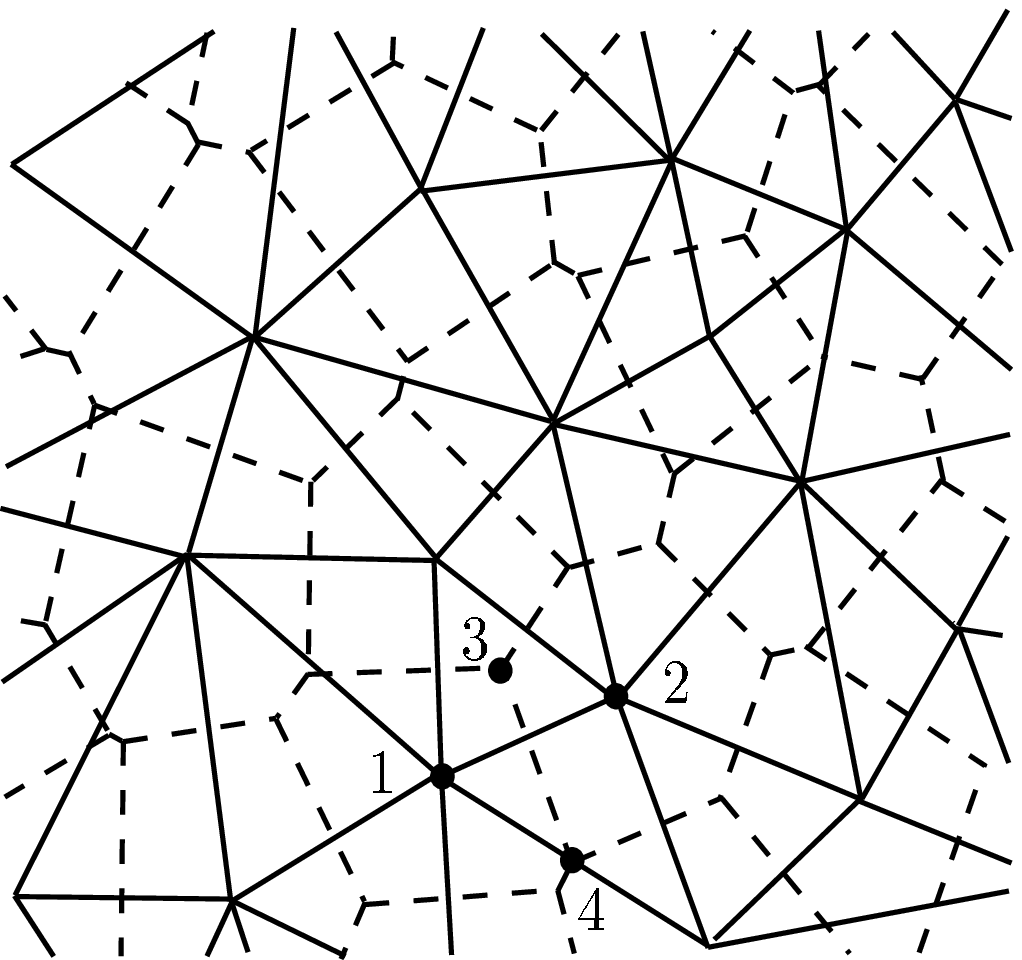,width=6cm} & \epsfig{file=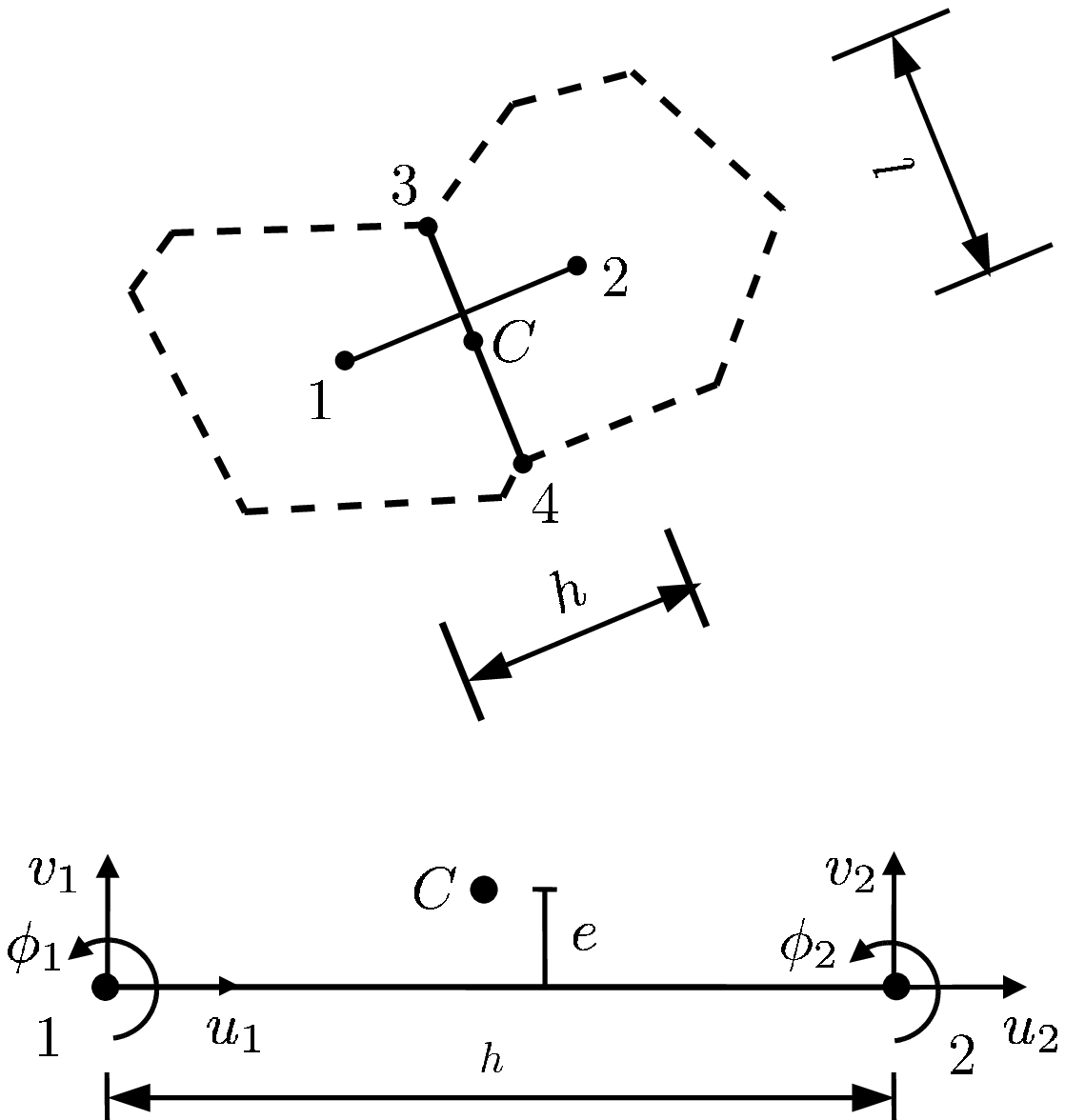,width=6cm} \\
(a) & (b)
\end{tabular}
\end{center}
\caption{Discretisation: (a) Lattice elements (solid lines) and cross-sections (dashed lines) obtained from the Delaunay triangulation and dual Voronoi tessellation, respectively. (b) Degrees of freedom $u_1$, $v_1$, $\phi_1$, $u_2$, $v_2$ and $\phi_2$ of the lattice element of length $h$ and cross-section width $l$ in the local coordinate system. The point $C$ at which the interface model is evaluated is in the center of the polygon facet at a distance $e$ from the center of the lattice element.}
\label{fig:vorDel}
\end{figure}

In the present work, shrinkage induced cracking was described by means of a lattice approach \cite{BolSai98} combined with a damage-plasticity constitutive law \cite{GraRem08}.
Here, the modelling approach is briefly reviewed.
Nodes are placed randomly in the specimen constrained by a minimum distance $d_{\rm m}$, i.e. the smaller $d_{\rm m}$ is, the smaller the average element length (Fig.~\ref{fig:vorDel}a).
Based on these randomly placed nodes, the spatial arrangement of lattice elements is determined by a Voronoi tesselation.
The cross-sections of the lattice elements, which connect the nodes (Voronoi sides), are the edges of the Voronoi polygons (Fig.~\ref{fig:vorDel}a).
Each node has three degrees of freedom, two translations and one rotation, shown in the local coordinate system in Figure~\ref{fig:vorDel}b.
The degrees of freedom $\mathbf{u}_{\rm e} = \left\{u_1, v_1, \phi_1, u_2, v_2, \phi_2\right\}^{\rm T}$ of two nodes of a lattice element are related to the displacement discontinuities $\mathbf{u}_{\rm c} = \left\{u_{\rm c}, v_{\rm c}\right\}^{\rm T}$ at the mid-point $C$ of the cross-section by
\begin{equation}\label{eq:uc}
\mathbf{u}_{\rm c} = \mathbf{B} \mathbf{u}_{\rm e}
\end{equation}
where
\begin{equation}\label{eq:bMatrix}
\mathbf{B} = \begin{bmatrix}
-1 & 0 & e & 1 & 0 & -e\\
0 & -1 & -h/2 & 0 & 1 & -h/2
\end{bmatrix}  
\end{equation}
The variable $h$ is the element length and $e$ is the eccentricity, defined as the distance between the mid-points of the lattice element and the corresponding cross-section, respectively (Figure~\ref{fig:vorDel}b).  
The element stiffness in the local co-ordinate system is
\begin{equation}
\mathbf{K}_{\rm e} = \dfrac{l}{h} \mathbf{B}^{\rm T} \mathbf{D} \mathbf{B}
\end{equation}
where $l$ is the length of the cross-section (Figure~\ref{fig:vorDel}b) and $\mathbf{D}$ is the constitutive matrix relating stresses to strains.
Different material phases are modelled by placing nodes strategically, so that cross-sections of lattice elements form the boundary between the phases \cite{YipMohBol05}.

The constitutive model of the lattice elements is a combination of plasticity and damage mechanics \cite{GraRem08}, which relates stresses to strains.
The strains $\boldsymbol{\varepsilon} = \left(\varepsilon_{\rm n}, \varepsilon_{\rm s}\right)^T$ are determined from the displacement jump $\mathbf{u}_{\rm c} = \left(u_{\rm n}, u_{\rm s}\right)^T$ at mid-point $C$ in Eq.~(\ref{eq:uc}) as
\begin{equation}
\boldsymbol{\varepsilon} = \dfrac{\mathbf{u}_{\rm c}}{h}
\end{equation}
These strains are related to the nominal stress $\boldsymbol{\sigma} = \left(\sigma_{\rm n}, \sigma_{\rm s}\right)^T$ as 
\begin{equation} \label{eq:totStressStrain}
\boldsymbol{\sigma} = \left(1-\omega \right) \mathbf{D}_{\rm e} \left(\boldsymbol{\varepsilon} - \boldsymbol{\varepsilon}_{\rm p} - \boldsymbol{\varepsilon}_{\rm s}\right) = \left(1-\omega\right) \bar{\boldsymbol{\sigma}}
\end{equation}
where $\omega$ is the damage variable, $\mathbf{D}_{\rm e}$ is the elastic stiffness,  $\boldsymbol{\varepsilon}_{\rm p} = \left(\varepsilon_{\rm pn}, \varepsilon_{\rm ps}\right)^T$ is the plastic strain, $\boldsymbol{\varepsilon}_{\rm s} = \left(\varepsilon_{\rm s}, 0\right)^T$ is the shrinkage eigenstrain and  $\bar{\boldsymbol{\sigma}} = \left(\bar{\sigma}_{\rm n}, \bar{\sigma}_{\rm s}\right)^T$ is the effective stress.
The elastic stiffness is 
\begin{equation}
\mathbf{D}_{\rm e} = \begin{Bmatrix} E & 0\\
  0 & \gamma E
\end{Bmatrix}
\end{equation}
where $E$ and $\gamma$ are model parameters controlling both the Young's modulus and Poisson's ratio of the material \cite{GriMus01}.
For a regular lattice and plane stress, Poisson's ratio $\nu$ is
\begin{equation}
\nu = \dfrac{1-\gamma}{3+\gamma}
\end{equation}
The plasticity part of the model is based on the effective stress $\bar{\boldsymbol{\sigma}}$ and consists of an elliptic yield surface, an associated flow rule, an evolution law for the hardening parameter and loading and unloading conditions.
A detailed description of the components of the plasticity model is presented in \cite{GraRem08}.
The initial yield surface of the plasticity model is determined by the tensile strength $f_{\rm t}$, the shear strength $s f_{\rm t}$ and the compressive strength $c f_{\rm t}$.
The evolution of the yield surface during hardening is controlled by the model parameter $\mu$, which is defined as the ratio of permanent and reversible inelastic displacements.
The damage part is formulated so that linear stress inelastic displacement laws for pure tension and compression are obtained, which are characterised by the fracture energies $G_{\rm ft}$ and $G_{\rm fc}$, respectively. 

\begin{figure} [t]
\begin{center}
\epsfig{file=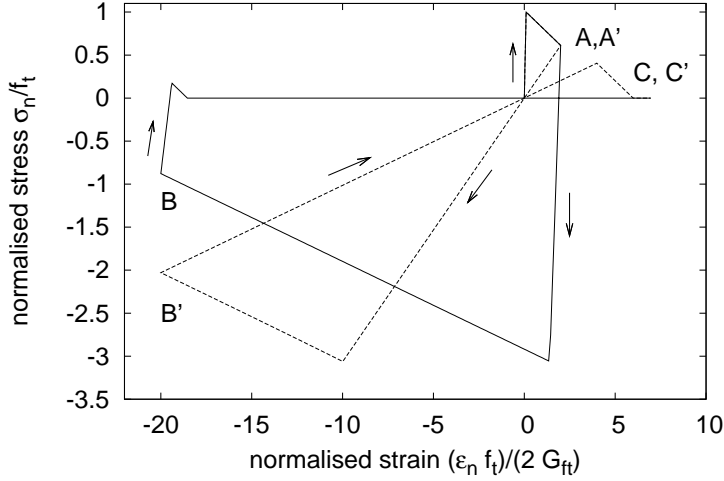, width=10cm}
\end{center}
\caption{Stress-strain response for fluctuating normal strains for  $\mu = 1$ (solid line) and $\mu = 0$ (dashed line).}
\label{fig:constCyclic}
\end{figure}

The constitutive response of the interface model is demonstrated in Fig.~\ref{fig:constCyclic} by the stress-strain response for fluctuating normal strains for $\mu = 1$ and $\mu = 0$.
The normal strain is increased to point $A$ ($A^{'}$). 
Then the strain is reduced to point $B$ ($B^{'}$) and again increased to point $C$ ($C^{'}$).
For $\mu \rightarrow 0$, a pure damage-mechanics response is obtained, which is characterised by reversible inelastic strains.
For $\mu = 1$, on the other hand, a pure plasticity model is obtained.
The unloading is elastic and the compressive strength is reached sooner than for $\mu = 0$.
The equivalent crack opening $\tilde{w}_{\rm c}$ is defined by the equivalent inelastic displacement, which for the present damage-plasticity model is defined as 
\begin{equation} \label{eq:equivCrack}
\tilde{w}_{\rm c} = \|\mathbf{w}_{\rm c}\|
\end{equation}
where
\begin{equation}
\mathbf{w}_{\rm c} = h \left(\boldsymbol{\varepsilon}_{\rm p} + \omega \left(\boldsymbol{\varepsilon} - \boldsymbol{\varepsilon}_{\rm p}\right)\right)
\end{equation}
The inelastic displacement vector $\mathbf{w}_{\rm c}$ is composed of a permanent and reversible part, defined as $h \boldsymbol{\varepsilon}_{\rm p}$ and $h \omega \left(\boldsymbol{\varepsilon} - \boldsymbol{\varepsilon}_{\rm p}\right)$, respectively.

\section{Nonlinear finite element analysis of shrinkage induced micro-cracking}
\begin{figure}
\begin{center}
\begin{tabular}{cc}
\epsfig{file=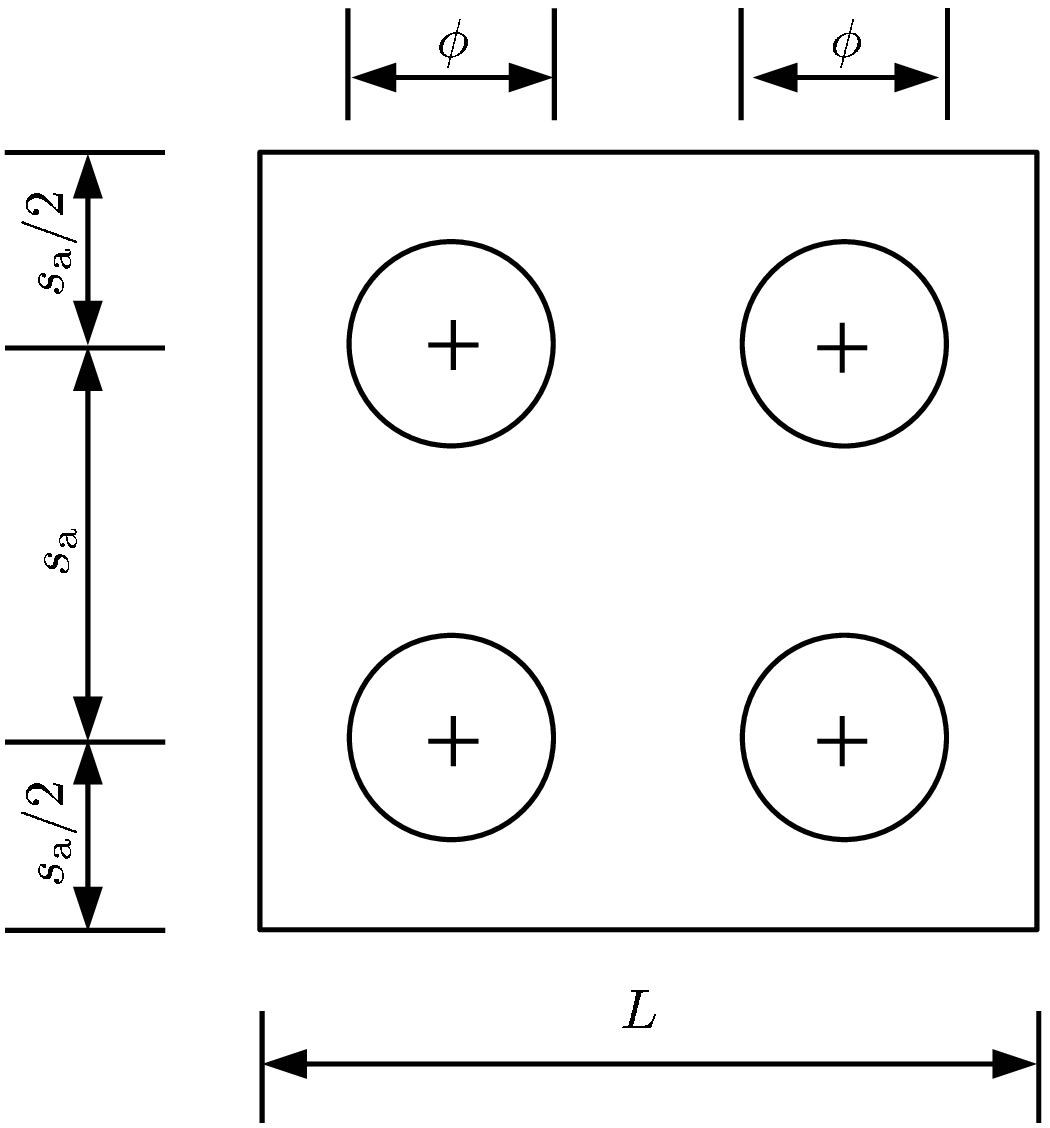, width=6cm} & \epsfig{file=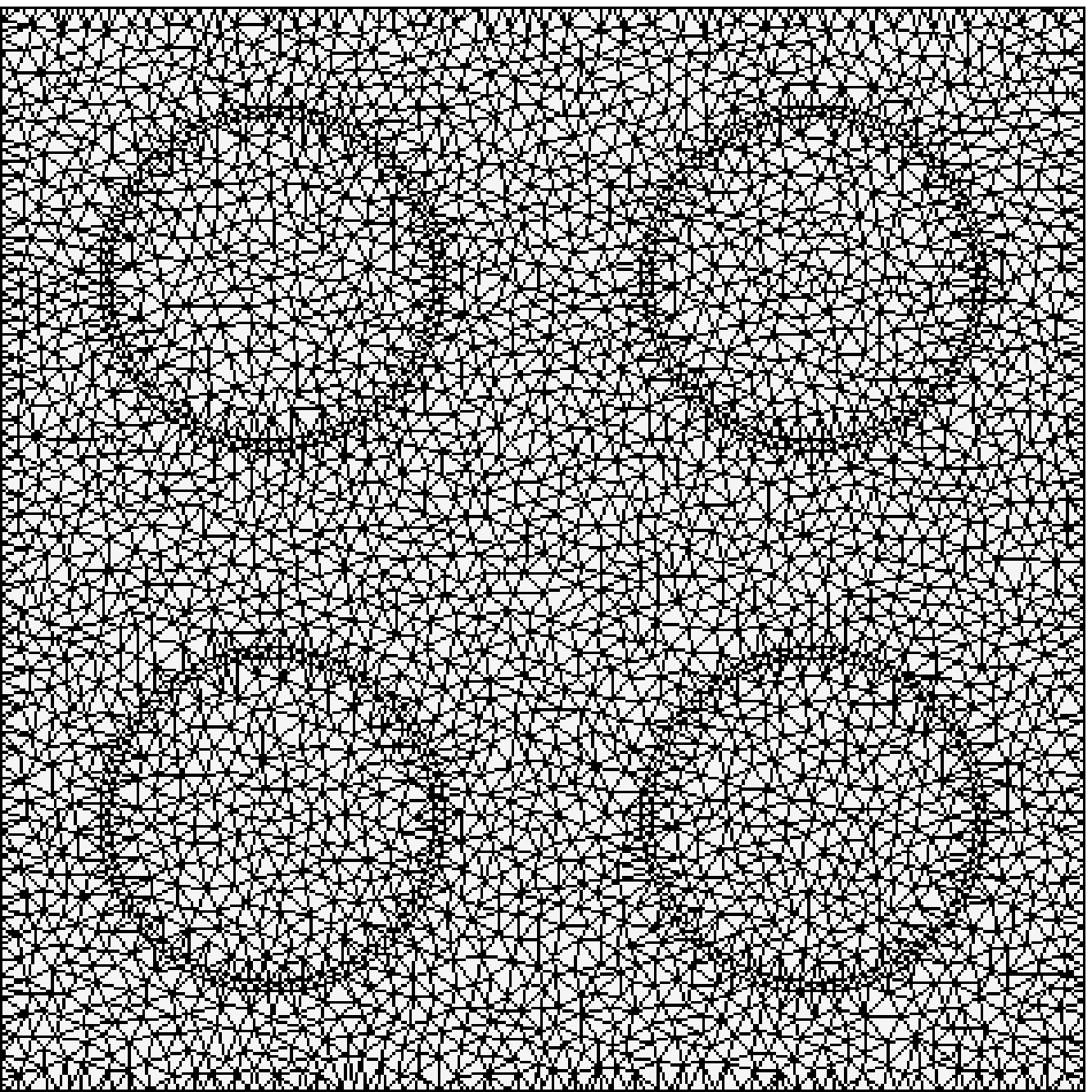, width=4.5cm}\\
(a) & (b)
\end{tabular}
\end{center}
\caption{(a) Geometry of the test specimen of length $L$. Aggregate size $\phi$ and volume ratio $\rho$ are varied. Specimen length $L$ is determined according to Eq.~(\ref{eq:length}). (b) Lattice for numerical analysis for $\rho = 0.3$ and $\phi = 16$~mm.}
\label{fig:geometry}
\end{figure}

Shrinkage induced micro-cracking was analysed by means of the nonlinear finite element approach described above.
The elements representing the cement paste were subjected to an incrementally applied uniform shrinkage strain up to $\varepsilon_{\rm s} = 0.5$~$\%$ (Eq.~\ref{eq:totStressStrain}).
This value was chosen for the simulation to represent a relatively severe shrinkage of the neat cement paste on first-drying to low humidities. 
For example, Helmuth and Turk \cite{HelTur67} reported first-drying shrinkage values of 0.35-0.70\% for neat cement pastes with w/c ratios of 0.3-0.6, cured for 6-30 months and dried at 47\% r.h. Further drying to 7\% r.h. produced additional irreversible shrinkage of 0.1\%. 
In another study, Fu et al. \cite{FuGuXie94} measured shrinkage eigenstrain values of 0.2-1.3\% by oven drying neat cement pastes (w/c ratios 0.25-0.70, cured for 6 months) at increasing temperature up to 110$^\circ$C. 
The influence of aggregate volume fraction and aggregate diameter was studied. Aggregate volume fractions $\rho = 0.5, 0.3$~and~0.1 were modelled. 
Furthermore, four different aggregate diameters ($\phi = 16, 8, 4,$~and~$2$~mm) were used. 
The geometry of the specimen analysed for all volume fractions and aggregate sizes is shown in Figure~\ref{fig:geometry}a.
The length $L$ of the specimen in Fig.~\ref{fig:geometry}a is 
\begin{equation}\label{eq:length}
L = \sqrt{\dfrac{\pi \phi^2}{\rho}}
\end{equation}
Thus, the smaller the aggregate diameter $\phi$ and the greater the volume ratio $\rho$, the smaller is the specimen length $L$.
The smallest separation distance between aggregate particles ($s_{\rm a}$-$\phi$), i.e. minimum width of the cement paste decreases with increasing aggregate volume fraction at constant aggregate size, as would be expected in the case of real mortars with increasing aggregate fraction. 
At constant aggregate fraction, the separation between aggregate particles increases with increasing aggregate size, similar to the case of a mortar compared to concrete at the same aggregate content. 
The ratio of the aggregate diameter $\phi$ and the minimum distance $d_{\rm m}$ for the nodes of the lattice was chosen as 64/3.
We will show later that varying this ratio has little effect on the simulation results.
Thus, the same detail of discretisation was applied, independently of the size of aggregates. 
For a diameter of $\phi=4$~mm for instance, the minimum distance corresponds to $d_{\rm m} = 0.1875$~mm.
The material parameters for the constitutive model were chosen according to Tab.~\ref{tab:param} \cite{Gra08}.
\begin{table}[t]
\caption{Model parameters.}
\label{tab:param}
\vspace{6pt} \center
\begin{tabular}{ccccccccc}
  Phase & $E$ [GPa] & $\gamma$ & $f_{\rm t}$ [MPa] & $s$ & $c$ & $G_{\rm ft}$ [J/mm$^2$] & $G_{\rm fc}$ [J/mm$^2$] & $\mu$ \\\hline
  Cement paste & 40 & 0.33 & 6.5 & 2 & 20 & 100 & 100000 & 0 \\
  ITZ & 57.1 & 0.33 & 3.25 & 2 & 20 & 50 & 50000 & 0 \\
  Aggregate & 100 & 0.33 & - & - & - & - & - & -
\end{tabular}
\end{table}
Aggregates were modelled elastically. 
Lattice elements crossing the boundary between aggregates and cement paste represent the average response of the interfacial transition zones and the two adjacent material phases.
In all the analyses, the length of the lattice elements is significantly greater than the width of the highly non-uniform interfacial transition zones, which is usually in the range of 10s of $\mu$m \cite{ScrCruLau04}.
Therefore, the stiffness of these lattice elements was approximated as 
\begin{equation}
E_{\rm ITZ} = \left(\dfrac{1}{2E_{\rm a}} + \dfrac{1}{2E_{\rm m}}\right)^{-1}
\end{equation}
where $E_{\rm a}$ and $E_{\rm m}$ are the Young's modulus of aggregate and mortar respectively.
Consequently, the Young's modulus of these elements is independent of the element size.
The strength of these lattice elements is determined by the strength of the ITZ, which is the weakest link.
Here, the strength and fracture energy ratio of cement paste and ITZ was chosen as 2.  
This ratio is an approximate value for samples with a relatively weak ITZ. For instance, Hsu and Slate \cite{HsuSla63} observed that the average tensile bond strength of the paste-aggregate interface varied from 50-75\% of the paste tensile strength, depending on the aggregate type, surface roughness and w/c ratio (0.265-0.36). The strength ratio, however, was observed to be independent of curing age (3-90 days).

\begin{figure}
\begin{center}
\begin{tabular}{cc}
\epsfig{file=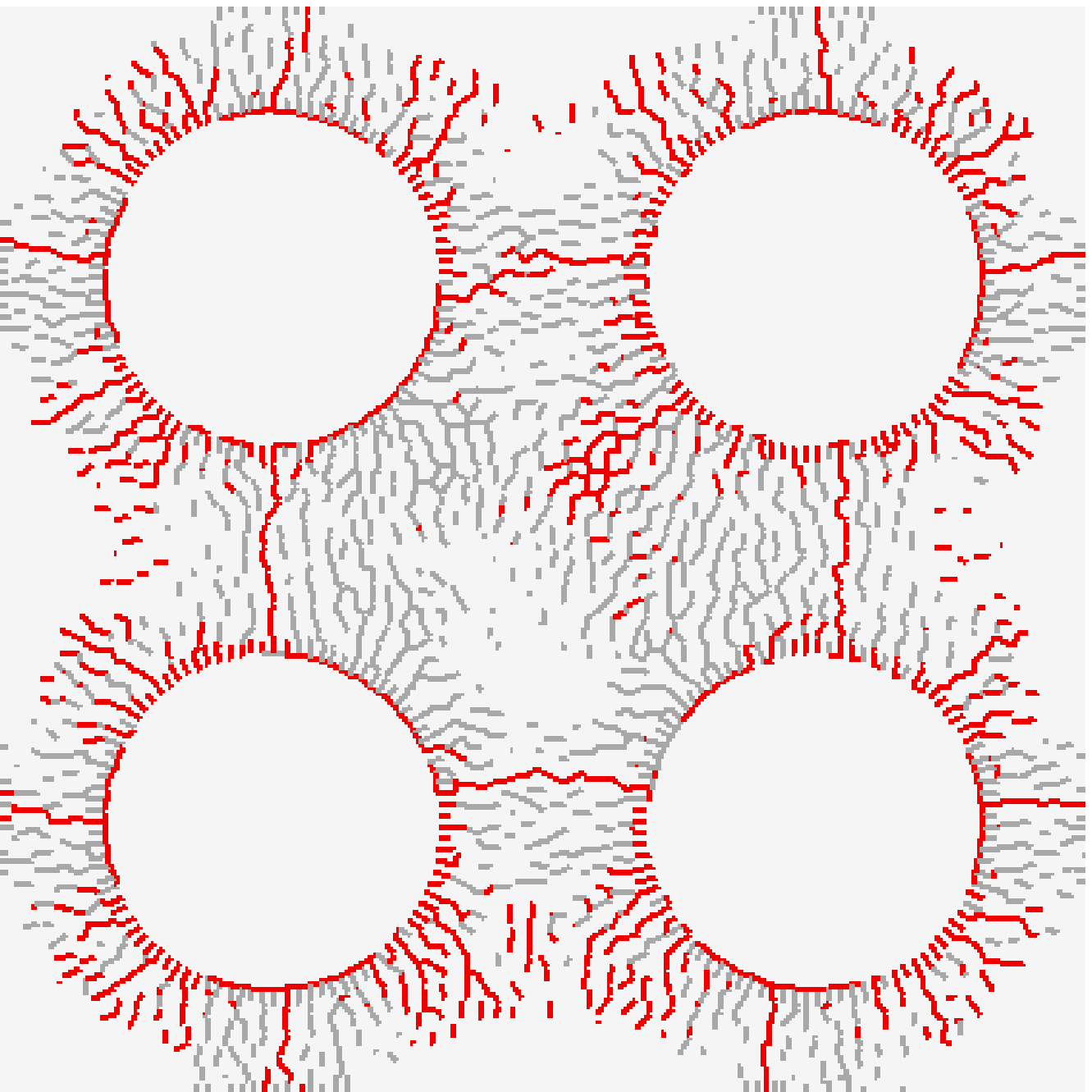, width=5cm} & \epsfig{file=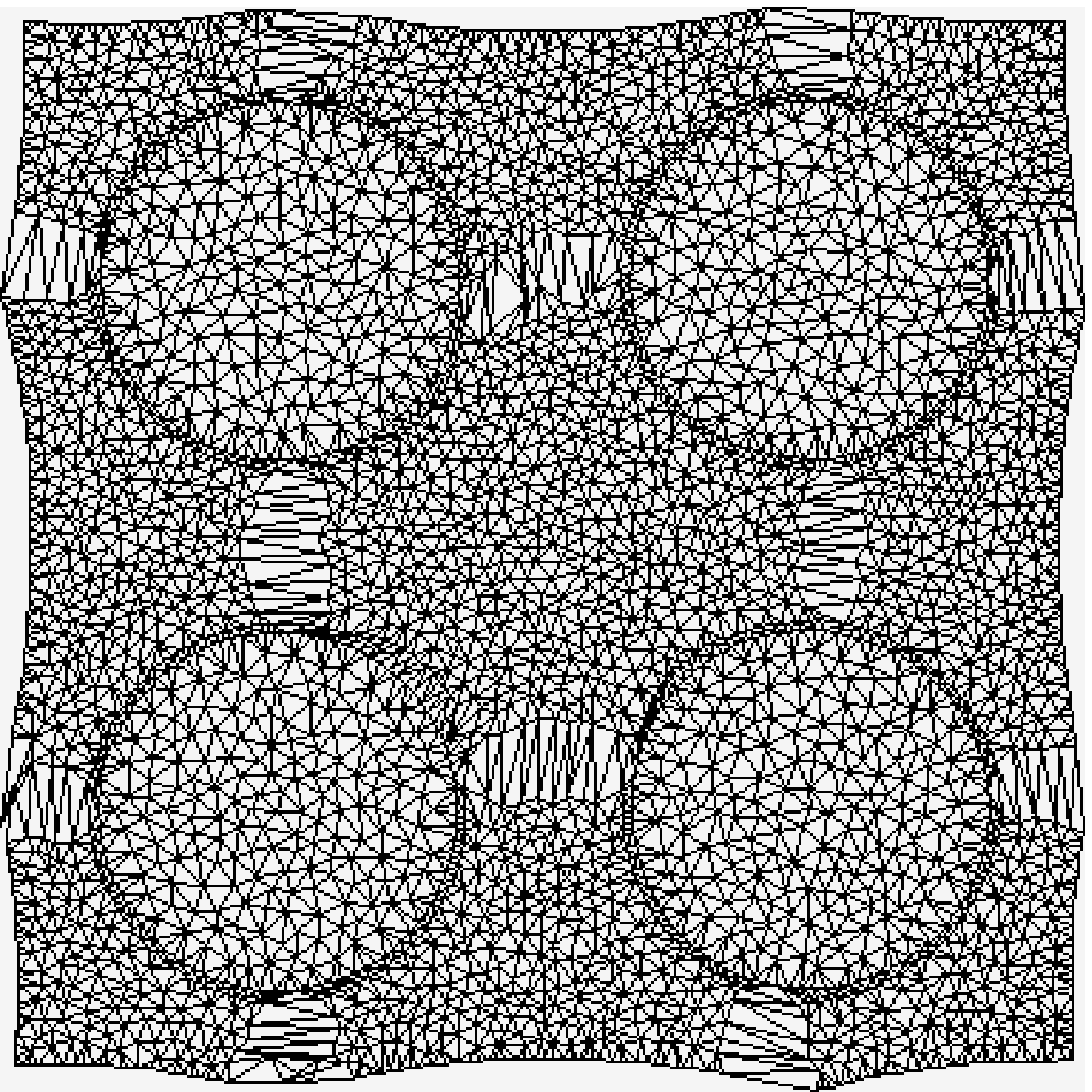, width=5cm}\\
(a) & (b)
\end{tabular}
\end{center}
\caption{(a) Crack patterns obtained for $\rho = 0.3$ and $\phi = 16$~mm at a shrinkage strain in the matrix of $\varepsilon_{\rm s} = 0.5$~$\%$. Black lines indicate active cracks, which increase at this stage of analysis. Grey lines represent inactive cracks, which do not increase at this stage of analysis. (b) Deformations (magnified by a factor of 50) at the same shrinkage strain.}
\label{fig:crack}
\end{figure}

With $\mu=0.001$, the constitutive model response is close to a perfect elasto-damage response (Fig.~\ref{fig:constCyclic}), which is characterised by the absence of permanent displacements.
For an aggregate diameter of $\phi=16$~mm and a volume fraction of $\rho = 0.3$, the crack patterns and the deformed mesh are shown in Fig.~\ref{fig:crack}a~and~b.
Black lines mark cross-sections of lattice elements, for which the maximum crack width is reached at this stage of the analysis.
This crack stage is defined as active.
On the other hand, grey lines represent lattice elements, which reached their maximum crack widths at an earlier stage of the analysis. 
The crack opening in these elements is either reducing, or increasing again after an earlier reduction.
The simulated widths of the active and passive cracks range between 0~and~70~$\mu$m.
%Cracks occur both in the ITZ and matrix.
Some bond cracks between the aggregate and paste can be observed, however most cracks appear to originate at the aggregate surface and propagate towards the matrix.
With the deformed mesh in Fig.~\ref{fig:crack}b, it is  illustrated that the deformations are localised in a few cracks, which connect the aggregates in a regular square pattern. 
Qualitatively very similar crack patterns were obtained for the other aggregate diameters and volume fractions.
This pattern of cracking seems consistent to that observed by Hsu \cite{Hsu63} in '2D' model samples made of sandstone discs arranged in a square grid and filled with paste that is subsequently subjected to drying shrinkage. 
Depending on the separation between the aggregates, cracks were seen to occur at the interface (i.e. bond cracks), near the shortest distance and diagonally at the largest distance between aggregates. 

\begin{figure}
\begin{center}
\begin{tabular}{cc}
\epsfig{file=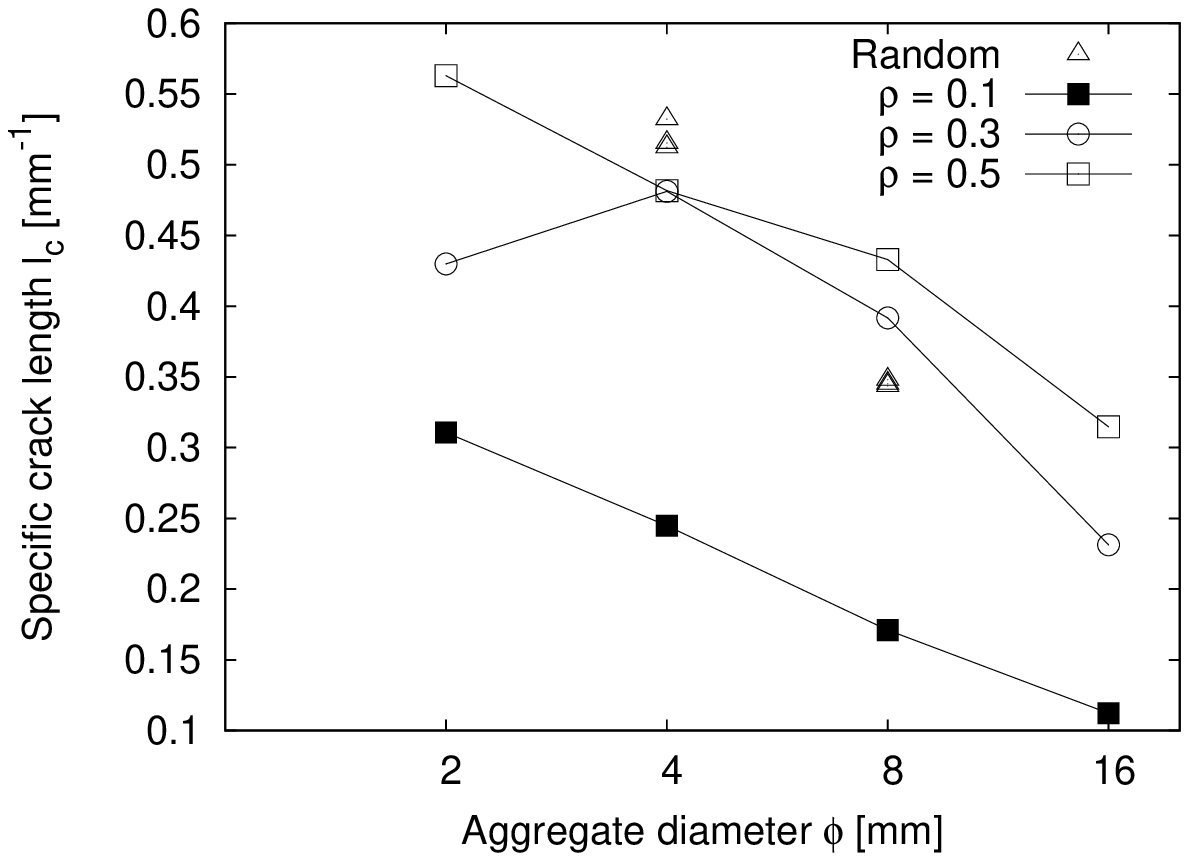, width=8cm} & \epsfig{file=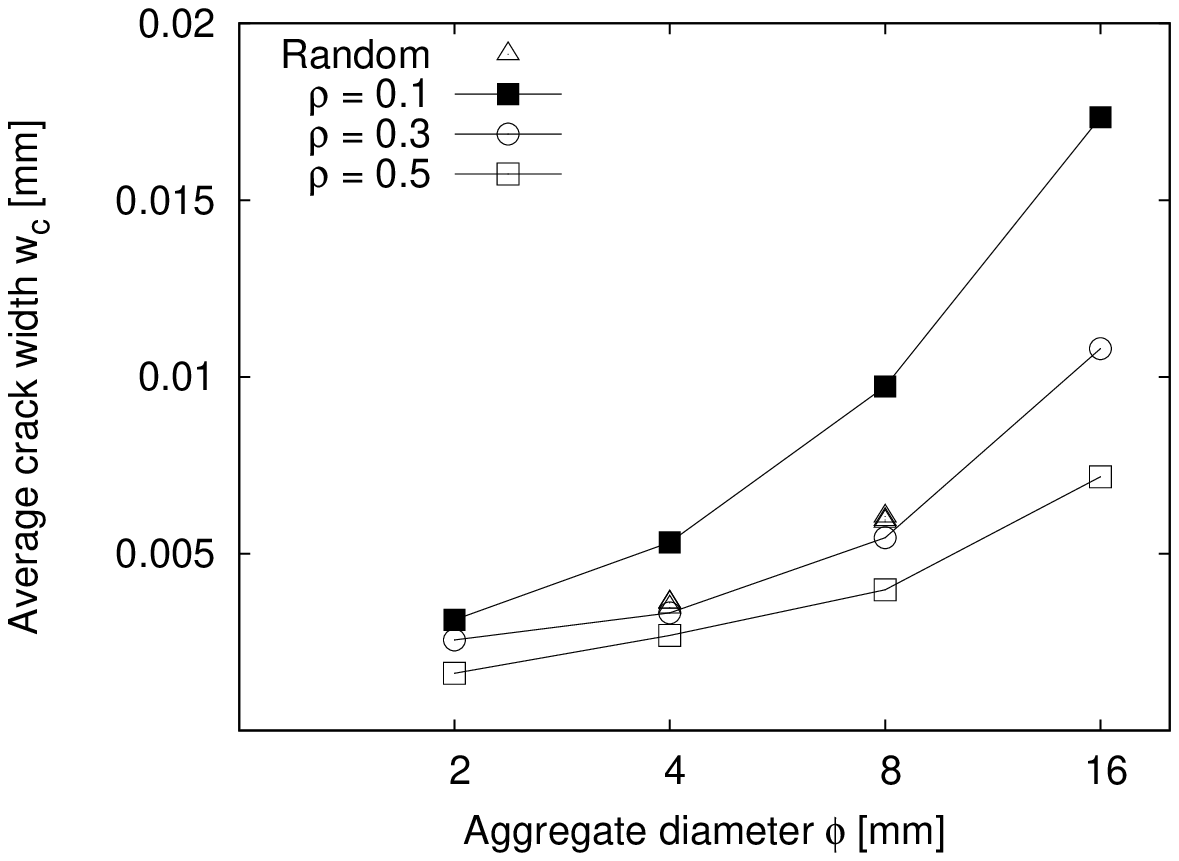, width=8cm}\\
(a) & (b)
\end{tabular}
\end{center}
\caption{(a) Specific crack length $l_{\rm c}$ in Eq.~(\protect\ref{eq:crackLength}) versus aggregate diameter $\phi$ for volume fractions of $\rho = 0.1, 0.3$ and $0.5$ at a shrinkage strain of $\varepsilon_{\rm s} = 0.5$~$\%$. (b) Average crack width $\bar{w}_{\rm c}$ in Eq.~(\protect\ref{eq:crackWidth}) versus aggregate diameter $\phi$ for volume fractions of $\rho = 0.1, 0.3$ and $0.5$ at a shrinkage strain of $\varepsilon_{\rm s} = 0.5$~$\%$.}
\label{fig:length}
\end{figure}

Length and width of micro-cracks for different volume fractions and aggregate diameters were compared.
The specific crack length $l_{\rm c}$ is
\begin{equation} \label{eq:crackLength}
l_{\rm c} = \dfrac{1}{A} \sum_{i = 1}^{n_{\rm c}} l_i
\end{equation}
where $A=L^2$ is the area of the specimen, $l_i$ is the width of the cross-section (Fig~\ref{fig:vorDel}) of lattice element $i$, and $n_{\rm c}$ is the number of cracked lattice elements.
Furthermore, the average crack width is
\begin{equation} \label{eq:crackWidth}
\bar{w}_{\rm c} = \dfrac{1}{l_{\rm c}A} \sum_{i = 1}^{n_{\rm c}} l_i \tilde{w}_{\rm {c} i} =  \dfrac{\sum_{i = 1}^{n_{\rm c}} l_i \tilde{w}_{\rm {c} i}}{\sum_{i = 1}^{n_{\rm c}} l_i}
\end{equation}
where $\tilde{w}_{\rm ci}$ is the crack width of element $i$ (Eq.~\ref{eq:equivCrack}).
The influence of the aggregate diameter and volume fraction on the specific crack length and average crack width are shown in Figs.~\ref{fig:length}a~and~b, respectively.
The specific crack length in Fig.~\ref{fig:length}a increases with decreasing aggregate size at constant volume fraction. 
An increase of volume fraction at constant aggregate size results in a decrease of the specific crack length.
Furthermore, the average crack width (Fig.~\ref{fig:length}b) decreases with decreasing aggregate size at constant volume fraction. 
Again, an increase of the volume fraction at constant aggregate size results in a decrease of the crack width.
Consequently, a decrease of the aggregate size results in more micro-cracking, which is in agreement with experimental observations made in \cite{WonZobBue08}.
The crack width is controlled by the spacing of the dominant cracks (Fig.~\ref{fig:crack}b), which is a function of aggregate diameter and volume fraction.
The greater the spacing between cracks, the greater is the average crack width.

\begin{figure}
\begin{center}
\begin{tabular}{cc}
\epsfig{file=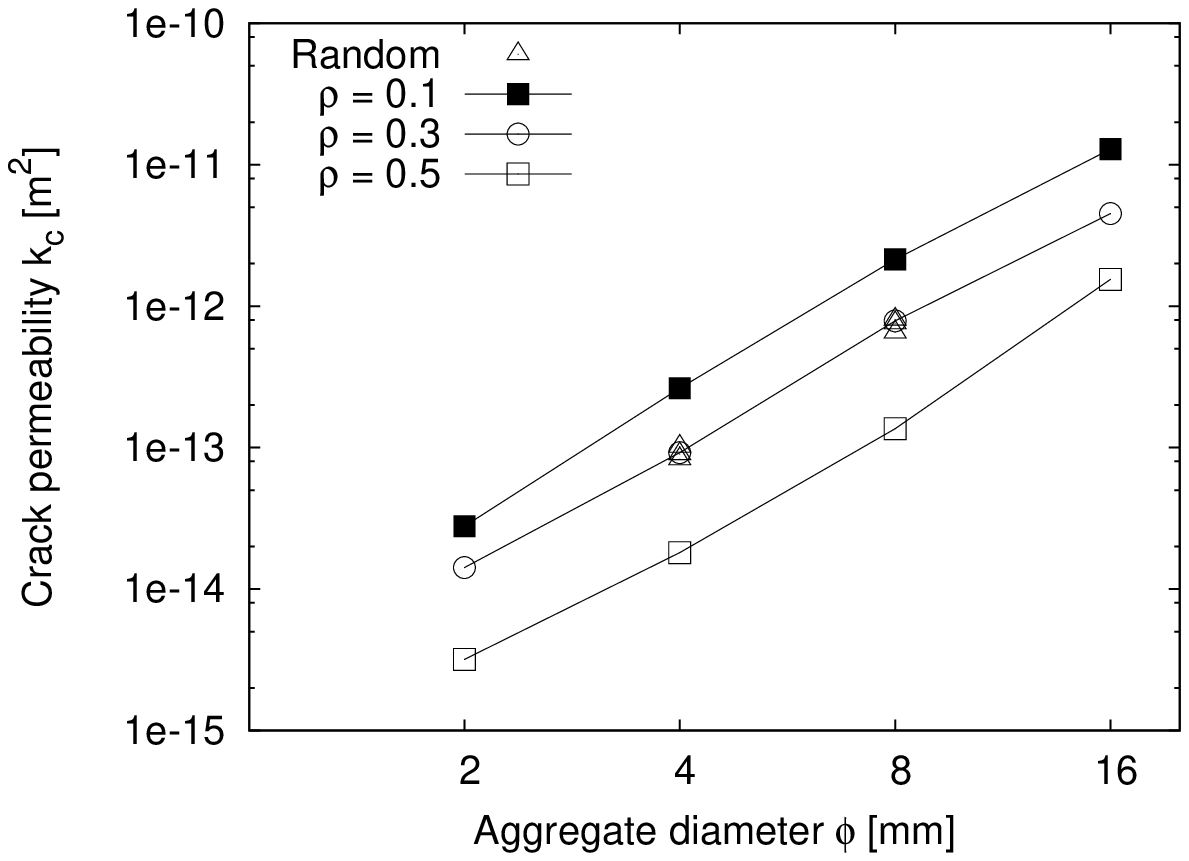, width=8cm} & \epsfig{file=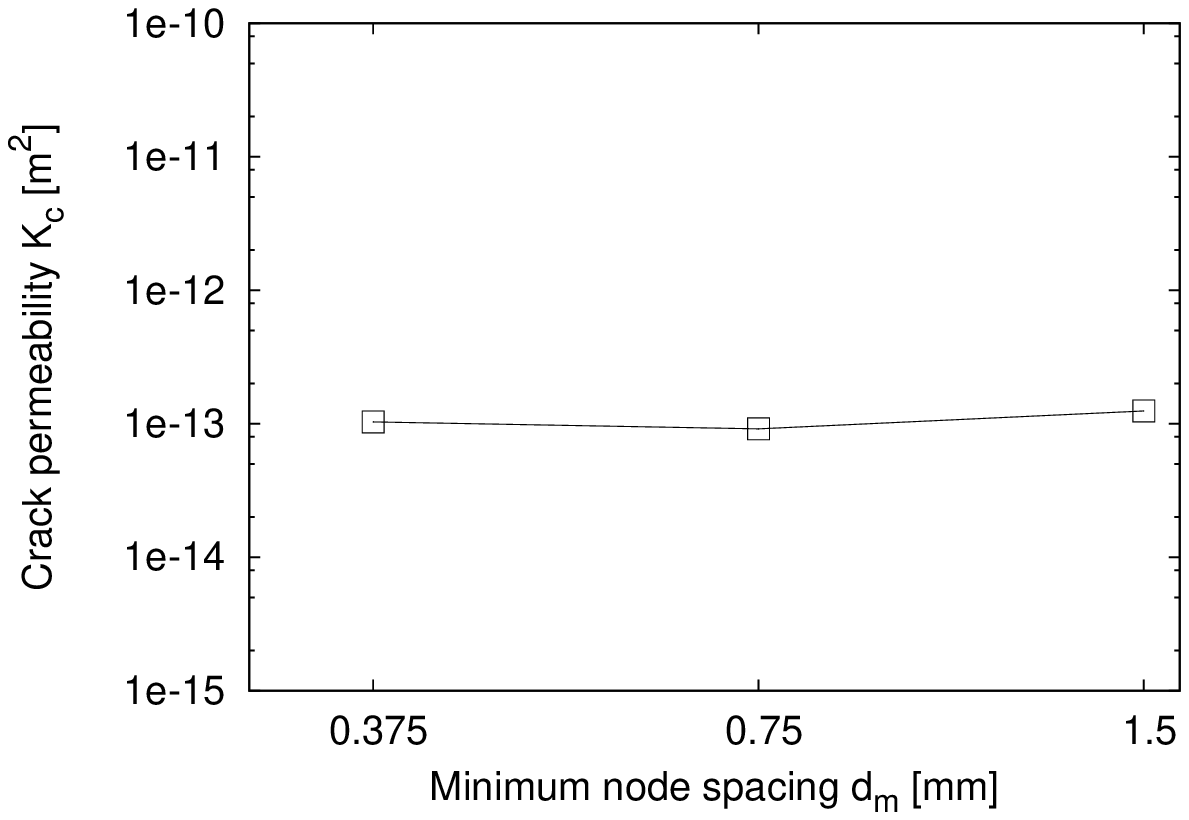, width=8cm}\\
(a) & (b)
\end{tabular}
\end{center}
\caption{(a) Crack permeability $k_{\rm c}$ (Eq.~\protect\ref{eq:crackPerm}) versus aggregate diameter $\phi$ for volume fractions of $\rho = 0.1, 0.3$ and $0.5$ at a shrinkage strain of $\varepsilon_{\rm s} = 0.5$~$\%$. (b) Minimum spacing of lattice nodes versus permeability $k_{\rm c}$ for $\rho = 0.3$ and $\phi = 4$~mm.}
\label{fig:con}
\end{figure}

Crack width is closely related to transport properties of concrete, in particular in the case of flow under a pressure gradient.
Assuming that the paste matrix is dense so that flow occurs predominantely through the cracks, the permeability $k_{\rm c}$ in the out-of-plane direction due to cracking can be described by the cube of the crack width as
\begin{equation} \label{eq:crackPerm}
k_{\rm c} = \dfrac{\xi}{A} \sum_{i = 1}^{n_{\rm c}} l_{i} \tilde{w}^3_{\rm{c} i}
\end{equation}
where $\xi$ is a material constant \cite{WitWanIwaGal80, AldGhaSha00}.
The influence of aggregate diameter and volume fraction on $k_{\rm c}$ is shown in Fig.~\ref{fig:con}a on a log-log scale for $\xi=1$.
Note that what is of interest here is not the actual value of the estimated permeability, but the change in permeability caused by varying either the aggregate volume fraction or particle size.
At a constant aggregate volume fraction, increasing the aggregate diameter from 2 to 16~mm caused approximately a 2.5 orders of magnitude increase in permeability. At constant aggregate diameter, increasing the aggregate volume fraction from 0.1 to 0.5 produced about 1 order of magnitude decrease in permeability.
The aggregate diameter influences the permeability strongly, since the crack width, which increases with increasing aggregate diameter (Fig.~\ref{fig:length}b), enters Eq.~(\ref{eq:crackPerm}) in its cube.

In Fig.~\ref{fig:crack}a many inactive cracks (grey lines) are visible, whereas the displacements are localised in only a few zones as illustrated by the deformed specimen in Fig.~\ref{fig:crack}b. 
These inactive cracks have small widths ($<$1~$\mu$m), but the evaluation of crack length in Eq.~(\ref{eq:crackLength}) and crack width in Eq.~(\ref{eq:crackWidth}) would be strongly influenced by this initial cracking.
To avoid misinterpretation, only lattice elements with crack-openings greater than 1~$\mu$m were considered in the evaluation of Eqs.~(\ref{eq:crackLength})~to~(\ref{eq:crackPerm}).

\section{Discussion}
The edge length $L$ of the specimen (Eq.~\ref{eq:length}) depends on the aggregate diameter $\phi$ and the volume fraction $\rho$.
Thus, the specimen area for which crack length, crack opening and hydraulic conductivity were evaluated, differed depending on the micro-structure considered.
Comparison of the results for the different micro-structures is only valid if the average properties are independent of the specimen area.
Here it is demonstrated that this is the case.
It is assumed that the square crack pattern, shown in Fig.~\ref{fig:crack}b in the form of cracks connecting the aggregate particles with a spacing $s_{\rm a}$, is independent of the number of aggregate particles considered.
Thus, the number of dominant cracks in one direction of the specimen is $n_{\rm c} = L/s_{\rm a}$. For instance, for the crack patterns shown in Fig.~\ref{fig:crack}b, $n_{\rm c} = 2$, whereas for a specimen of double the size it would be 4, since $s_{\rm a}$ is related to the material structure and is independent of the specimen length $L$.
With this assumption, the specific crack length $l_{\rm c}$ in Eq.~(\ref{eq:crackLength}), which is defined as the total length of cracks divided by the specimen area, is
\begin{equation}
l_{\rm c} = \dfrac{2 n_{\rm c} \left(L-n_{\rm c} \phi\right)}{A} = \dfrac{L^2 \dfrac{2}{s_{\rm a}} \left(1-\dfrac{2\phi}{s_{\rm a}} \right)}{L^2} = \dfrac{2}{s_{\rm a}} \left(1- \dfrac{2\phi}{s_{\rm a}}\right)
\end{equation}
where $L-n_{\rm c} \phi$ is the length of one crack.
The specific crack length $l_{\rm c}$ depends only on the parameters $s_{\rm a}$ and $\phi$, but not on the specimen length $L$. 
An increase of $L$, at constant $\phi$ and $s_{\rm a}$, does not change $l_{\rm c}$ if the influence of the specimen boundaries, which is studied later, is negligible. 
Equivalent observations can be made for the average crack width $\bar{w}_{\rm c}$ and $k_{\rm c}$.

\begin{figure}
\begin{center}
\begin{tabular}{cc}
\epsfig{file=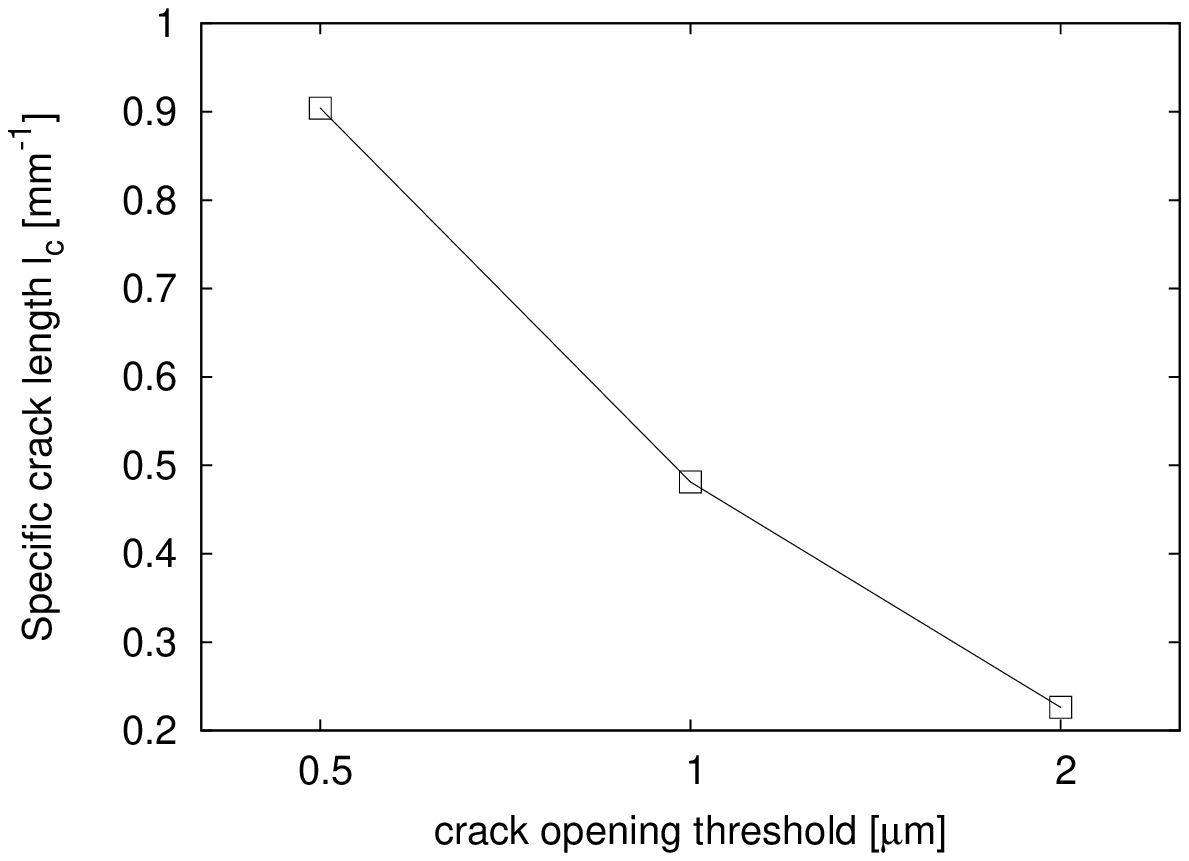, width=8cm} & \epsfig{file=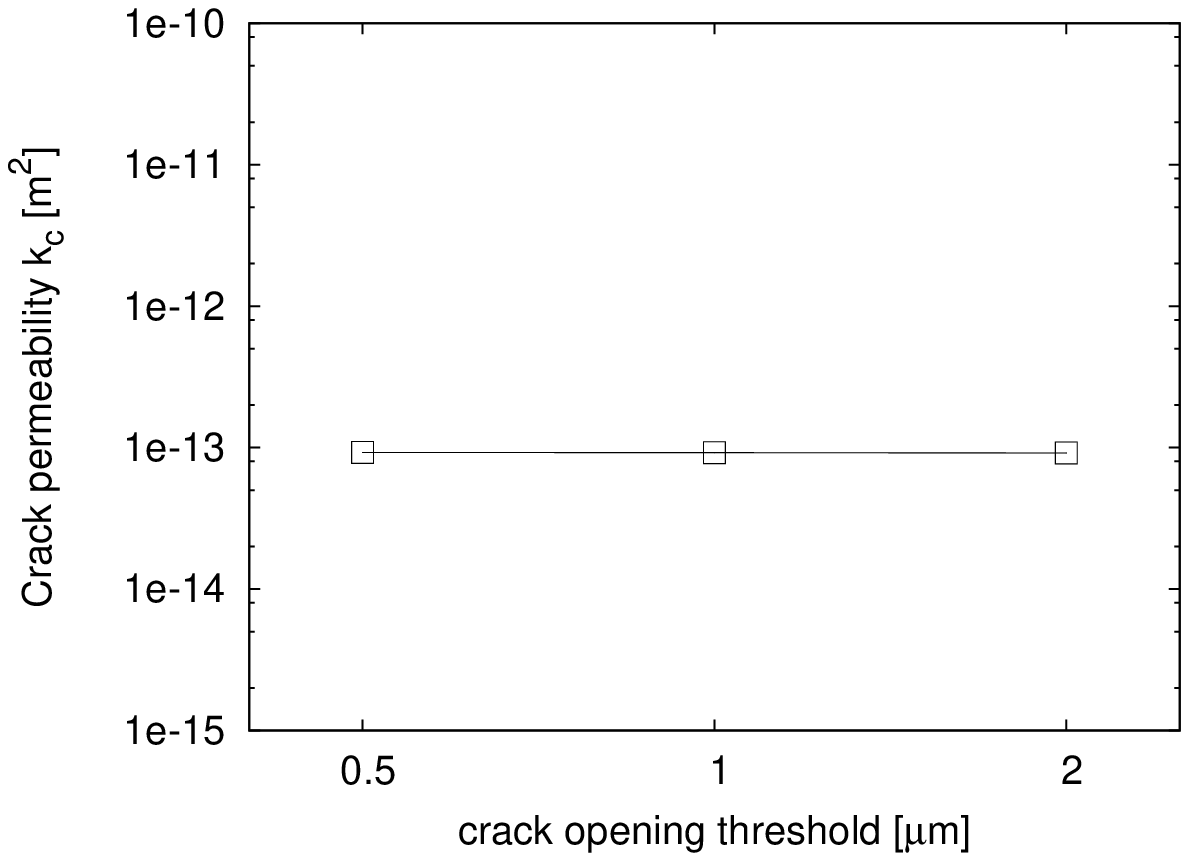, width=8cm}\\
(a) & (b)
\end{tabular}
\end{center}
\caption{Influence of crack opening threshold for $\rho = 0.3$ and $\phi = 4$~mm. (a) Specific crack length versus crack opening threshold. (b) Crack permeability $k_{\rm c}$ (Eq.~\protect\ref{eq:crackPerm}) versus crack opening threshold.}
\label{fig:init}
\end{figure}

The damage-plasticity constitutive model used for the finite element analyses was designed to result in mechanical responses, which are independent of the size of the lattice elements \cite{GraRem08}. 
In the present study, a mesh-size independent description of the crack opening is important, since the analysis for varying aggregate diameters for the aggregate arrangement in Fig.~\ref{fig:geometry}a was performed with the same diameter mesh size ratio of 64/3.
Consequently, the smaller the aggregate diameter, the smaller is the length of the lattice elements used. 
To investigate a possible influence of element size on crack-openings, two additional analyses with $\phi$ to $d_{\rm m}$ ratios of $32/3$ and $128/3$ were performed for a volume fraction $\rho = 0.3$ and a constant aggregate diameter $\phi=4$~mm.
The influence of element size on permeability (Eq.~\ref{eq:crackPerm}) with $\xi=1$ is shown in Fig.~\ref{fig:con}b on log-log scale.
The evaluation of permeability is almost independent of the element size.
The change in the simulated permeability due to element size is small and negligible compared to the change in permeability due to varying aggregate size and volume fraction (Fig.~\ref{fig:con}a).
%NEW:Start----------------------------------
As mentioned above, the results of the analyses were evaluated for lattice elements with crack openings greater than 1~$\mu$m.
This threshold was applied to avoid misinterpretation due to distributed initial cracking.
The influence of the crack opening threshold on crack length (Eq.~\ref{eq:crackLength}) and permeability (Eq.~\ref{eq:crackPerm}) with $\xi=1$ is shown in Fig.~\ref{fig:init}a~and~b, respectively.
The specific crack length is strongly influenced by the crack threshold. On the other hand, crack permeability is almost independent of the threshold, since lattice elements with small crack openings contribute little to the overall crack permeability.
%NEW:End---------------------------------------------

\begin{figure}
\begin{center}
\begin{tabular}{cc}
\epsfig{file=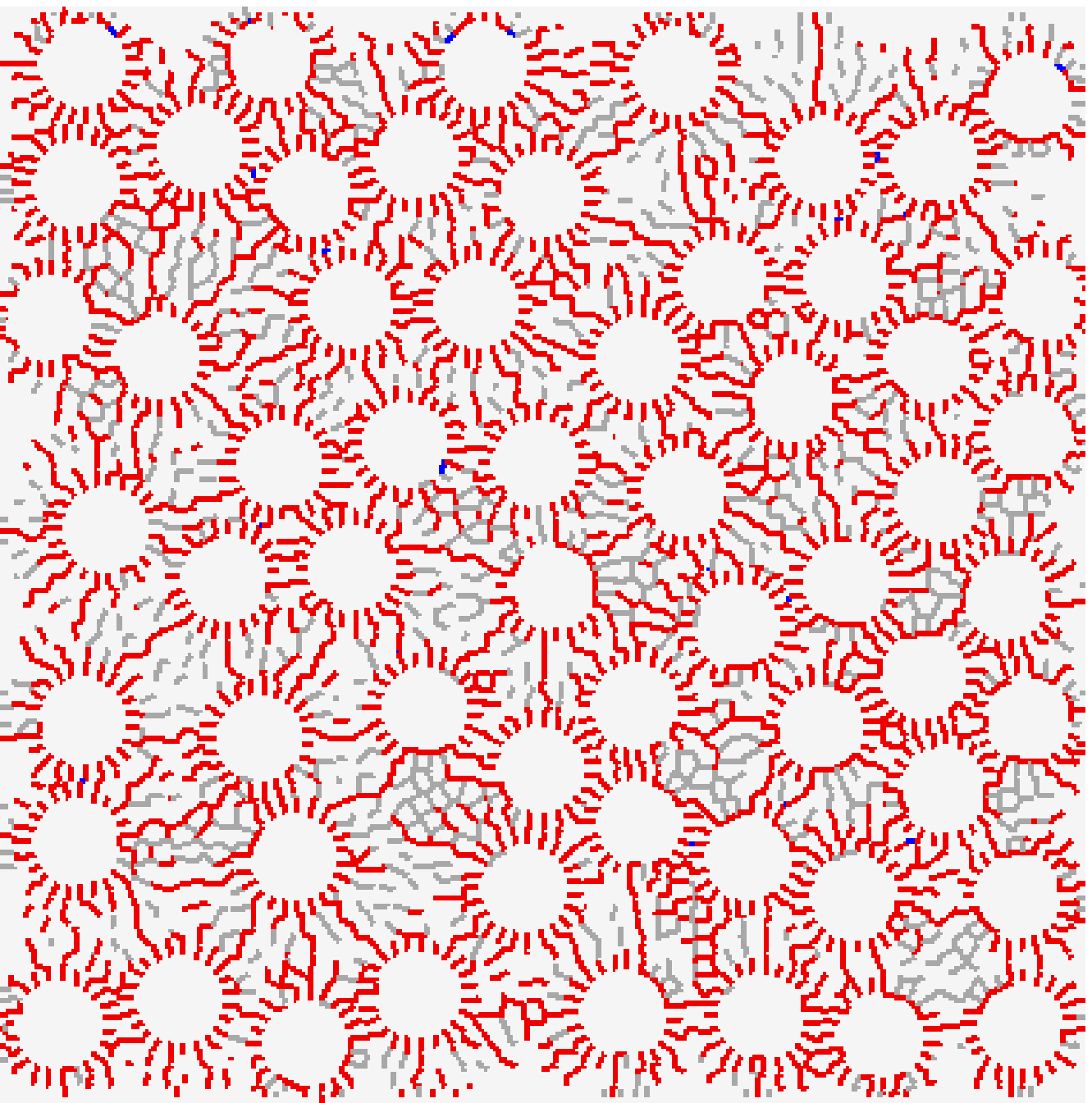, width=5cm} & \epsfig{file=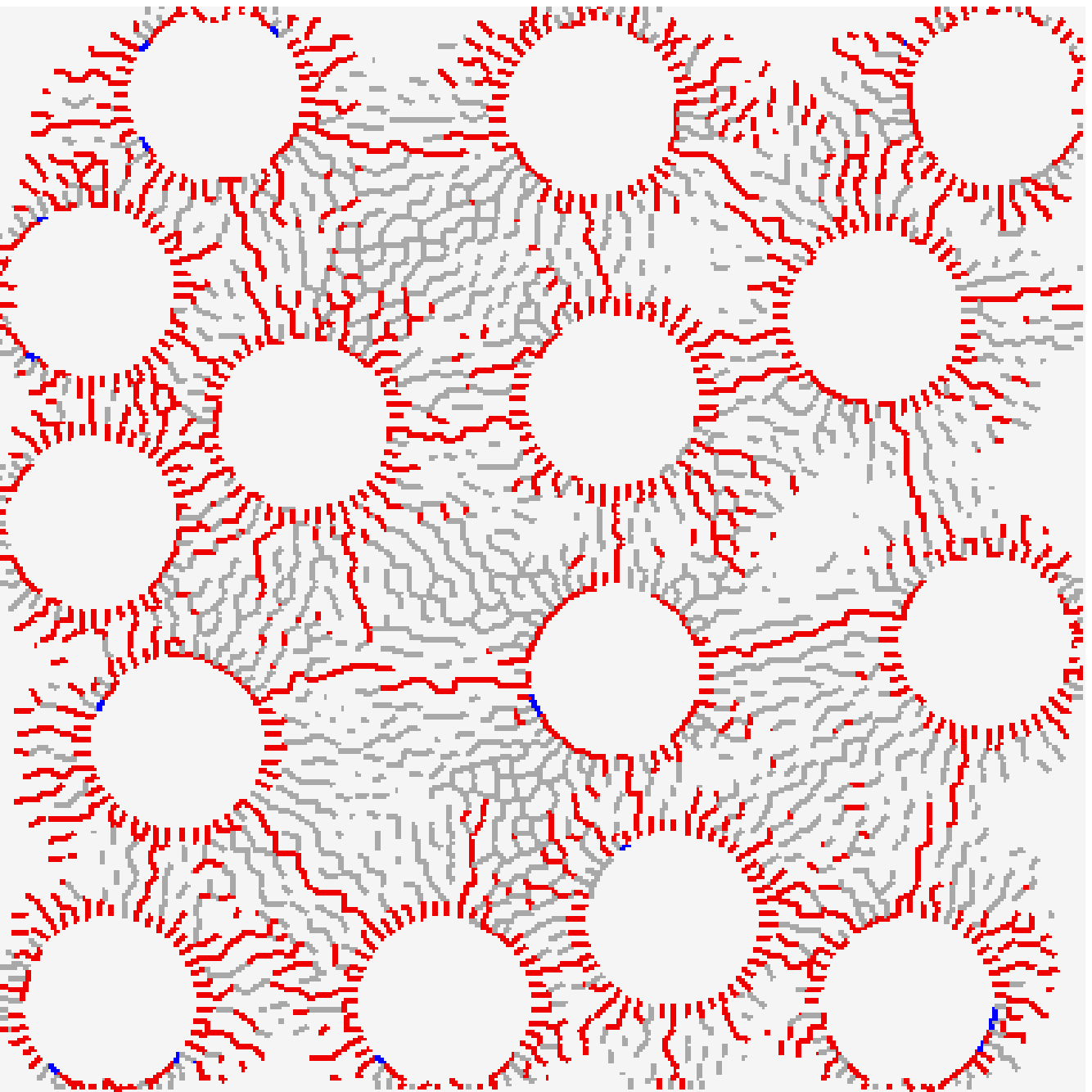, width=5cm}\\
(a) & (b)
\end{tabular}
\end{center}
\caption{Crack patterns for $\rho = 0.3$ for (a) $\phi = 4$~mm and (b) $\phi = 8$~mm at a shrinkage strain in the matrix of $\varepsilon_{\rm s} = 0.5$~$\%$. Black lines indicate active cracks, which increase at this stage of analysis. Grey lines represent inactive cracks, which do not increase at this stage of analysis.}
\label{fig:randomCrack}
\end{figure}

The meso-structure used in this study is a strong idealisation of concrete and results in regular square crack patterns (Fig.~\ref{fig:crack}b), which differ considerably from the random crack patterns observed in experiments \cite{WonZobBue08}. 
Therefore, additional analyses were carried out to investigate the influence of random aggregate arrangements for a volume fraction of $\rho=0.3$ and two aggregate diameters of $\phi = 4$~and~8~mm. 
The specimen length was chosen as $L=50$~mm.
Three analyses for each aggregate diameter were carried out to determine the mean of the random results.
The same material properties as for the regular analyses were used.
The crack patterns for two random analyses for aggregate diameters $\phi = 4$~and~8~mm are shown in Fig.~\ref{fig:randomCrack}a~and~b, respectively.
The specific crack length $l_{\rm c}$, the average crack width $\bar{w}_{\rm c}$ and the change of permeability $k_{\rm c}$ obtained by means of the random analyses are presented in Fig.~\ref{fig:length}a~and~b, and~Fig~\ref{fig:con}a (triangular symbols) together with the results obtained from the analysis with regular aggregate arrangements.
Although the resulting crack pattern differs considerably, as would be expected, the specific crack length, average crack width and permeability results for random and regular aggregate arrangements are in good agreement.
Consequently, the regular arrangement seems appropriate to investigate the influence of aggregate diameter on shrinkage induced cracking.

\begin{figure}
\begin{center}
\epsfig{file=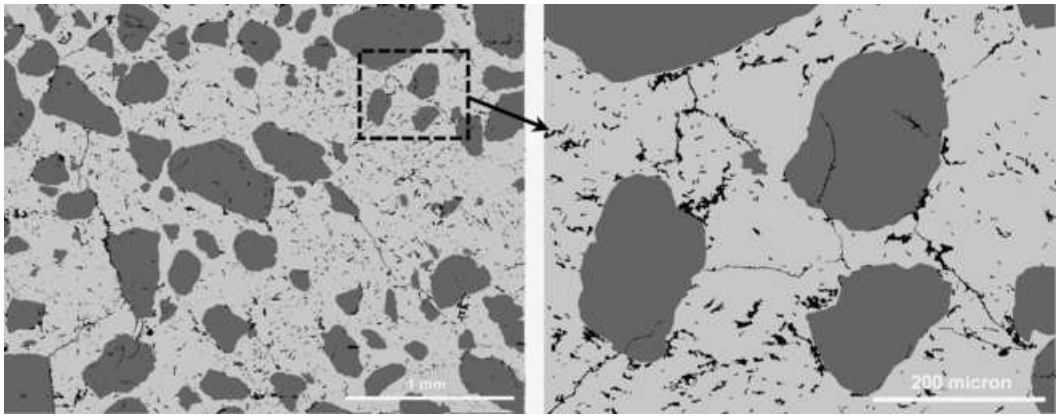, width=14cm}
\end{center}
\caption{Segmented BSE images of a 90-day mortar (w/c 0.5, 40\% vol. sand) that was conditioned at 50$^\circ$C, 10\%r.h. showing the microcracks, pores, aggregates and cement paste. The image on the left is a composte image produced by aligning 3$\times$3 images captured at $100\times$, to give a larger field of view (3100$\times$400$\mu$m). The image on the right was obtained at $200\times$. Pores with aspect ration $<2$ were filtered to highlight the microcracks.}
\label{fig:crackExp}
\end{figure}

It may be useful to relate the findings of this study to that of the experimental results by Wong et al. \cite{WonZobBue08}. 
A direct comparison is not possible because of the approximations used in this study, although certain trends seem to be captured by our simulations. 
Using image analysis, Wong et al. \cite{WonZobBue08} found that the concretes have more microcracks, and that the microcracks have lower specific lengths, compared to analogous mortars subjected to the same drying conditions. 
Fig.~\ref{fig:crackExp} shows crack patterns observed on a mortar containing 40\% vol. sand (0.5 w/c ratio, 90-day cured) after drying at 50$^\circ$C, 10\% r.h. using backscattered electron microscopy.
Examination of samples dried at 105$^\circ$C found more severe cracking, and significantly higher permeability \cite{WonZobBue08}. 
The microcracks have widths of about 0.5-10$\mu$m, although finer cracks were seen at higher magnifications. 
The crack pattern bears some resemblance to the simulations shown in Fig.~\ref{fig:randomCrack}, such as its random orientation, the partial bond cracks appearing around some aggregate particles and the matrix cracks that propagate through the paste and very often bridging several aggregate particles.
However, the simulations did not produce cracks that propagate through the aggregates.
Also, the simulated cracking appears to be more severe than that observed on the mortars dried at 50$^\circ$, 10\% r.h.
A possible reason for this is a mismatch in the model parameters used for the simulation.

Wong et al. \cite{WonZobBue08} found that increasing aggregate fraction of the mortar from 10\% to 50\% resulted in a decrease in measured permeability by about an order of magnitude (Fig.~\ref{fig:exp}b). 
Given the same drying treatment, concretes (with max. aggregate size 12.7~mm) have significantly higher permeability, about 1.5 orders of magnitude, than mortars (max. aggregate size 5~mm) at the same aggregate content. 
Interestingly, this change in permeability due to the aggregate content and size effect is close to the simulated values in Fig.~\ref{fig:con}. This obviously may be a fortuitous result since several important features were not captured by the model, for instance, the contribution of capillary pores, the effect of tortuosity and connectivity, shape and size distribution. Some may have an opposite effect or a minor contribution to transport properties, but clearly more work needs to be carried out. Nevertheless, the results of this study show that the aggregate content and particle size have a significant influence on the formation of microcracking induced by aggregate restrained shrinkage, which in turn affects the permeability of the composite.

\section{Conclusions}
In the present work the influence of aggregate size and volume fraction on shrinkage induced micro-cracking was studied numerically by means of the nonlinear finite element method. 
The work resulted in the following conclusions:
\begin{itemize}
\item The length of micro-cracks decreases with increasing aggregate diameter and increasing volume fraction.
\item The average crack width increases with increasing aggregate diameter and decreasing volume fraction.
\item The permeability, which is related to the cube of the crack width, increases with increasing aggregate diameter and decreasing volume fraction
\end{itemize}
Thus, the aggregate diameter and volume fraction influence the formation of microcracks induced by aggregate restrained shrinkage, which is in agreement with experimental observations in \cite{WonZobBue08} (Sec.~\ref{sec:experiments}).
However, the representation of aggregates as circular inclusions and the use of one aggregate size oversimplifies the micro-structure of concrete, which does not allow a direct comparison with experimental results.
In future work, the present modelling approach will be extended to 3D and the use of realistic aggregate grading curves will be included by applying multi-scale analysis approaches.

\section{Acknowledgements}
The simulations were performed with the object-oriented finite element package OOFEM \cite{Pat99,PatBit01} extended by the present authors. 
HSW and NRB acknowledge the financial support from the Engineering and Physical Sciences Research Council, UK. 

\bibliographystyle{plainnat}
\bibliography{general}

\clearpage

\end{document}